\documentclass[12pt]{amsart}
\newtheorem{prop}{Proposition}[section]
\newtheorem{coro}{Corollary}[section]

\newtheorem{exampleHLP}{Example}[section]

\newtheorem{remarkHLP}{Remark}[section]
\newenvironment{remark}[0]{\begin{remarkHLP}\rm}{\end{remarkHLP}}
\def\Em{{\mathcal M}}
\def\De{{\mathcal D}}

\def\Tr{{\rm Tr\,}}
\def\re{{\rm Re\,}}
\def\<{\langle}
\def\>{\rangle}
\begin{document}

\title{Geodesic distances on density matrices}
\author{Anna Jen\v cov\'a} 
\maketitle

\begin{center}
{\small Mathematical Institute, Slovak Academy of Sciences,\\
      \v Stef\'anikova 49 SK-814 73 Bratislava, Slovakia,\\
jenca@mat.savba.sk}
\end{center}

\vskip 1cm

\noindent {\bf Abstract.}{\small We find an upper bound for geodesic distances
associated to monotone Riemannian metrics on positive definite matrices and density matrices.}

\section{Introduction}

The notion and importance of Fisher  information is well established
in statistics and probability theory. As a measure of distinguishability 
of probability densities, the Fisher information was used by Rao to define a
Riemannian metric on probability spaces. 
On the simplex of probability vectors 
${\mathcal P}_n=\{p=(p_1,\dots,p_n),\ \sum_i p_i=1, p_i>0, i=1,\dots,n\}$, 
this is the unique metric contracting under markovian mappings, by the
Chentsov uniqueness theorem. On ${\mathcal P}_n$, the Fisher metric is
$$
\lambda_p(x,y)=\sum_ip_i^{-1}x_iy_i\qquad x,y\in T_p{\mathcal P}_n
$$
The geometry of ${\mathcal P}_n$ with this metric is quite simple. By
\begin{equation}\label{eq:sqrt}
p\mapsto 2(\sqrt{p_1},\dots,\sqrt{p_n}),
\end{equation}
it is isometric
with an open subset in the sphere of radius 2 in $\mathbb{R}^n$,
\cite{giis03}.
The metric can be  extended to the set $\Em_n=\{p=(p_1,\dots,p_n),\ 
p_i>0\}$
of all finite (strictly positive) measures on the 
set $\{1,\dots,n\}$. Using the isometry (\ref{eq:sqrt}) and elementary
geometry in $R^n$, we may compute the geodesic distance for the Fisher
metric in ${\mathcal P}_n$ and $\Em_n$:
$$
D(p,q)=2\arccos(\sum_i \sqrt {p_i}\sqrt{q_i})\qquad  p,q\in {\mathcal P}_n
$$
(the Bhattacharya distance) and
$$
d(p,q)=2(\sum_i(\sqrt{p_i}-\sqrt{q_i})^2)^{1/2}\qquad p,q\in\Em_n
$$
The last expression is related to the Hellinger distance
$H(p,q)$
by $d(p,q)=\sqrt{2H(p,q)}$. The Hellinger distance belongs
to the family of Czisz\'ar's $f$-divergences
$$
D_f(p,q)=\int f(q/p)dp
$$
here $f$ is a convex function. As it was shown in \cite{amari}, the metric
given by the hessian of 
$f$-divergence is a constant multiple of the Fisher  metric.

In the case of a quantum system, the situation becomes more complicated. In the
simplest case, the states of the system are represented by density matrices.
In analogy with manifolds of classical probability densities, a
quantum version of the Fisher information metric must be decreasing under
stochastic maps. 
Contrary to the classical case, this
monotonicity condition does not specify the metric uniquely.
In fact, it was shown by Petz that the monotone metrics can be labelled by
operator-monotone functions.

As it was mentioned in \cite{giis03}, there is no general formula for geodesic
path and distance for a general monotone metrics. Explicit expressions are
known only in two particular cases, namely the Bures metric and the 
Wigner-Yanase metric. In the present paper, we find an upper bound for the
geodesic
 distances for all monotone metrics. This is done in a simple way:
Following Uhlmann \cite{uhlm93,uhlm95}, we obtain the Bures geodesics from 
certain purifying lifts of curves of density matrices 
and then make use 
of a duality relation between the smallest (Bures) and the largest (RLD)
of monotone metrics. It is also shown that this upper bound is related to a
particular non-commutative version of the Hellinger distance.

\section{The manifold and monotone metrics.}

Let $M_n$ be the algebra of $n$ by $n$ complex matrices.  The set of faithful
positive linear functionals on $M_n$ is identified with the cone of positive
definite matrices.  This set, with the differentiable manifold
structure inherited from $M_n$, will be denoted by $\Em$. Let $\De\subset \Em$
denote the submanifold of density matrices in $\Em$, that is
$$
\De=\{ \rho\in \Em:\ \Tr \rho=1\}
$$
The tangent space to $\Em$ at $\rho\in \Em$ is $T_{\rho}\Em=\{ x\in M_n:\
x=x^*\}$.
If $\rho\in \De$, then the tangent space $T_{\rho}\De$ is the subspace of
traceless matrices in $T_{\rho}\Em$.

Let $\lambda$ be a Riemannian metric on $\Em$. Then we will say that $\lambda$ 
is a monotone metric if
$$
{\lambda}_{T(\rho)}(T(h),T(h))\le \lambda_{\rho}(h,h),
\qquad \rho\in \Em,\ h\in T_{\rho}\Em
$$
for all completely positive trace preserving maps $T$.
It is an important
result of Petz \cite{petz96} that a Riemannian metric is  monotone if and 
only if it has the form
$$
\lambda_{\rho}(h,k)=\Tr hJ_{\rho}(k)
$$
where $J_{\rho}$ is given by the operator mean
\begin{equation}\label{eq:monotone}
J_{\rho}=R_{\rho}^{-1}[f(L_{\rho}/R_{\rho})]^{-1}
\end{equation}

Here $L_{\rho}$ and $R_{\rho}$ are the left and the right
multiplication operator and $f:(0,\infty)\to \mathbb{R}$ is an
operator monotone  function which is symmetric, that is, $f(t)=tf(t^{-1})$.
It is immediate from (\ref{eq:monotone}) that under the normalization $f(1)=1$,
any monotone metric is equal to the Fisher metric on commutative submanifolds.
Moreover, we have
$$
\frac {2t}{1+t}\le f(t)\le \frac {1+t}2
$$
for  all  symmetric normalized operator monotone functions \cite{kuban}.  
Accordingly, 
there is a greatest and a smallest element in the set of monotone metrics.

The smallest monotone metric is obtained for  $f(t)=(1+t)/2$.
It is called the Bures metric,
because it is related to the Bures distance, see also Section
\ref{sec:geodesic}.
The operator
$$
J_{\rho}(h)=g,\qquad \rho g+g\rho=2h  
$$
is the symmetric logarithmic derivative, see \cite{holevo,uhlm93, brca}.

The greatest monotone metric corresponds to the function $f(t)=2t/(1+t)$. 
In this case $J_{\rho}$ is the right logarithmic derivative (RLD)
$$
J_{\rho}(h)= \frac 12 (\rho^{-1}h+h\rho^{-1})
$$
see \cite{holevo,petz96,peru}. More examples of monotone metrics can be found in
Section \ref{sec:wyd}.

\section{Standard representation and monotone metrics.}
\label{sec:representation}

The standard representation of the algebra $M_n$ is obtained if $M_n$ is
endowed with the Hilbert-Schmidt inner product
$$
\<x,y\>=\Tr x^*y
$$
Let us denote the resulting Hilbert space by $H$.
Then $M_n$ is represented on $H$ by
$$
\phi:\ M_n\to {\mathcal B}(H),\qquad a\mapsto L_a
$$
where $L_a$ is the left multiplication operator $L_aw=aw$, $w\in H$.  
Each element $\rho$  in $\Em$ has a vector representative, or purification,
$w$ in $H$, such that
$$
 \Tr \rho a=\<w,L_aw\>\qquad \forall a\in M_n
$$
Then $w\in H$ is a vector representative of $\rho\in \Em$ if and only if
$\rho=ww^*$. 

Let $\rho_t$, $t\in I$,
be a smooth curve in $\Em$. A curve $w_t$ in $H$, such that 
$w_t$ is a vector representative of $\rho_t$ for all $t\in I$ 
is called a lift of $\rho_t$. In this
case, the tangent vectors are related by
\begin{equation}\label{eq:projection}
\dot{\rho_t}=\dot w_tw^*_t+w_t\dot w^*_t
\end{equation}
Let us denote the corresponding projection of the tangent spaces
$T_wH\to T_{ww^*}\Em$ by $\Pi$.

Let $w_0w_0^*=\rho_0$. There are many lifts of $\rho_t$ through $w_0$.
Among such lifts, there is a unique lift with minimal 
Hilbert space length
$$
l_H(w_t)=\int_I \sqrt{\<\dot w_t,\dot w_t\>}dt.
$$
It will be called the horizontal lift.

The horizontal  lift was introduced
in \cite{uhlm93,uhlm95}, where  the 
geometric phase was extended to mixed states. 
It was shown that the above minimalization problem leads to the condition
\begin{equation}\label{eq:horizontal}
w_t^*\dot w_t=\dot w_t^*w_t
\end{equation}
for all $t$. The curves $w_t$ in $H$, satisfying this condition, are called
horizontal curves.
The tangent vectors to horizontal curves at $w\in H$
form a real vector subspace
$H_w=\{ gw,\ g=g^*\}$. Let $H_w$ be endowed with the inner product
$\re\<\cdot,\cdot\>$, then it is a real Hilbert space,
called the horizontal subspace. 
For each $h\in T_{ww^*}\Em$, there is a unique element $\hat h$ in $H_w$,
satisfying $h=\Pi(\hat h)$. It follows that the 
 inner product in $H_w$ can be projected onto $T_{ww^*}\Em$.
As it turns out, this projection defines a Riemannian  metric on $\Em$,
moreover
\begin{equation}\label{eq:metric}
4\re\<\hat h,\hat k\>=
2\Tr h(L_{\rho}+R_{\rho})^{-1}(k),\quad h,k\in T_{\rho}\Em
\end{equation}
is exactly the  Bures metric.

The commutant of $\phi(M_n)$ is the algebra of right multiplication operators 
$R_aw=wa$, $a\in M_n$, on $H$. For each $\sigma\in \Em$, there is an element  
$w\in H$,
such that
$$
\<w,R_aw\>=\Tr \sigma a
$$
This element is given by $\sigma=w^*w$. For each curve 
$\sigma_t$ in $\Em$, let us consider the curves $w_t$ in $H$ satisfying 
$w_t^*w_t=\sigma_t$. The tangent vectors of such curves satisfy 
$\dot{\sigma}_t=\tilde {\Pi}(\dot w_t)$, where
$\tilde{\Pi}:\ T_wH\to T_{w^*w}\Em$ is given by
$$
\tilde {\Pi}(x)=x^*w+w^*x 
$$
We may now proceed exactly  as before, choosing 
for each  $\sigma_t$ the shortest of these curves.  It is quite clear that
$w_t$ is the shortest curve  if and only if $w_t^*$ is  horizontal, 
equivalently, 
$\dot w_t\in \tilde H_{w_t}:=\{ w_tg,\ g=g^*\}$ 
for all $t$. Moreover,  we have 
\begin{equation}\label{eq:horiz}
x\in H_w\iff x^*\in \tilde H_{w^*}
\end{equation}
If we now 
project  the real Hilbert space structure  from $\tilde H_w$
to $T_{w^*w}\Em$,
using the  projection $\tilde {\Pi}$, we will, of course,
get the Bures metric again.  On the other hand, 
it is easy to see that for each $\rho=ww^*$ and  $h\in T_{\rho}\Em$, 
$\tilde h:=\frac 12h(w^*)^{-1}$  is the unique element in 
$\tilde H_w$ satisfying $h=\Pi(\tilde h)$.
We may therefore define
\begin{equation}\label{eq:RLDmetric}
\lambda_{\rho}(h,k):=4\re \<\tilde h,\tilde k\>=\frac 12 \Tr \rho^{-1}(hk+kh)
\end{equation}
which is the RLD metric. 
This shows that there is a duality relation between the Bures metric and
RLD, see also \cite{mats,ja02}.

\section{The geodesic distances}\label{sec:geodesic}

Let $\lambda$ be a Riemannian metric on $\Em$. A curve $\rho_t$, $t\in [0,1]$
is a geodesic path in $\Em$ if its length
$$
l_{\lambda}(\rho_t)=\int_0^1\sqrt{\lambda_{\rho_t}(\dot {\rho_t},\dot {\rho_t})}dt
$$
is the minimum of lengths of all curves connecting $\rho_0$ and $\rho_1$.
This  length is then the geodesic distance of $\rho_0$ and $\rho_1$.
Let us denote by $d_{\lambda}$ the geodesic distance for the metric 
$\lambda$ in
$\Em$ and by $D_{\lambda}$ the geodesic distance in $\De$.

For Bures metric, the geodesic paths and distances were obtained by
Uhlmann
\cite{uhlm93,uhlm95} as follows.
Let  $\rho_0$ and $\rho_1$ be two elements in
$\Em$ and let $\rho_t$ be a curve connecting them. If $w_t$ is the
horizontal lift of $\rho_t$, then by (\ref{eq:metric})
$$
l_{Bures}(\rho_t)=2l_H(w_t),
$$
hence minimizing the Bures length means minimizing the Hilbert space 
length of horizontal
lifts of curves connecting $\rho_0$ and $\rho_1$. From the definition of 
horizontality,  this  minimum is attained at the line segment 
$w_t=tw_1+(1-t)w_0$, such that  $\|w_0-w_1\|$ is minimal over 
$w_0w_0^*=\rho_0$, $w_1w_1^*=\rho_1$.  This happens if and only if  $w_1$ and
$w_0$ are parallel amplitudes, that is, these satisfy Uhlmann's parallelity
condition
\begin{equation}\label{eq:parallel}
w^*_1w_0\ge 0
\end{equation}
For each $w_0$ there is a unique  $w_1$ parallel to $w_0$,
given by \cite{uhlm95}
$$
w_1=\rho_0^{-1/2}(\rho_0^{1/2}\rho_1\rho_0^{1/2})^{1/2}\rho_0^{-1/2}w_0
$$
The  geodesic path in $\Em$, connecting $\rho_0$ and $\rho_1$ is then
$$
\rho_t=(tw_1+(1-t)w_0)(tw_1+(1-t)w_0)^*
$$
and the geodesic  distance is
$$
d_{Bures}(\rho_0,\rho_1)=2\|w_0-w_1\|=2\sqrt{\Tr
\rho_0+\Tr\rho_1-2\Tr(\rho_0^{1/2}\rho_1\rho_0^{1/2})^{1/2}}
$$
this is called the Bures distance. 

Let now $\rho_t$ be a curve in $\De$, then all lifts of $\rho_t$
are curves on the unit sphere $S$ in $H$. If $w_0,w_1\in S$, the shortest 
curve connecting them lies on the large circle in $S$ through them. The
length of such arcs for $w_0w_0^*=\rho_0$ and $w_1w_1^*=\rho_1$ is minimal 
if $w_0$ and $w_1$ are parallel amplitudes and, by definition,
in this case the arc is also horizontal. Hence, the Bures geodesic in $\De$ is
$$
\rho_t=\frac{(w_0+(1-t)w_1)(tw_0+(1-t)w_1)^*}
{\|tw_0+(1-t)w_1)\|^2}
$$
for parallel amplitudes $w_0$ and $w_1$ and the Bures distance
$$
D_{Bures}(\rho_0,\rho_1)=2\arccos \Tr w_0w_1^*=2\arccos{\Tr(\rho_0^{1/2}\rho_1\rho_0^{1/2})^{1/2}}
$$

The duality of the Bures and RLD metrics leads to the following upper bound 
for the RLD geodesic distance.
\begin{prop}\label{prop}
Let $\rho_0$, $\rho_1\in \Em$, then
$$
d_{RLD}(\rho_0,\rho_1)\le d_{Bures}(\rho_0,
\rho_0^{-1/2}(\rho_0\#\rho_1)^2\rho_0^{-1/2})
$$
where
$$
\rho_0\#\rho_1=\rho_0^{1/2}(\rho_0^{-1/2}\rho_1\rho_0^{-1/2})^{1/2}\rho_0^{1/2}
$$
is the geometric mean. If $\rho_0$ and $\rho_1$ are in $\De$, the same
holds for geodesic distances $D_{RLD}$ and $D_{Bures}$.
\end{prop}

\begin{proof} Let 
 $w_0=\rho_0^{1/2}$ and let  $w\in H$ be such
that $w_0$ and $w$ satisfy the parallelity condition (\ref{eq:parallel}).
Then the curve $w_t=tw+(1-t)w_0$ is the horizontal lift of the
Bures geodesic connecting $\rho_0$
and $ww^*$, in particular, $\dot w_t\in H_{w_t}$ for all $t$. 
Then $w_t^*$ is a lift of a curve $\rho_t$ in $\Em$, 
connecting $\rho_0$ and $w^*w$ and
by (\ref{eq:horiz}), $\dot w_t^*\in \tilde H_{w^*_t}$. 
Consequently, by (\ref{eq:RLDmetric}),
$$
d_{RLD}(\rho_0,w^*w)\le l_{RLD}(\rho_t)=2\|w^*-w_0^*\|=2\|w-w_0\|=
d_{Bures}(\rho_0,ww^*)
$$
From the parallelity condition, $w=qw_0$ for some $q=q^*>0$. Let us choose
$w$ such that
$$
\rho_1=w^*w=\rho_0^{1/2}q^2\rho_0^{1/2}
$$
then $q=(\rho_0^{-1/2}\rho_1\rho_0^{-1/2})^{1/2}$ and
$$
ww^*=\rho_0^{-1/2}(\rho_0\#\rho_1)^2\rho_0^{-1/2}
$$
The statement for  distances in $\De$ is proved exactly the same way.
\end{proof}

\begin{remark}
Let $w_0$, $w$ and $q$
be as in the proof of the previous Proposition, then we have
$$
\|w_0-w\|^2= \Tr \rho_0 +\Tr\rho_1-2\Tr \rho_0q
$$
and
\begin{equation}\label{eq:eq}
d_{Bures}(\rho_0,\rho_0^{-1/2}(\rho_1\#\rho_0)^2\rho_0^{-1/2})=
2\sqrt{\Tr\rho_0 +\Tr\rho_1-2\Tr\rho_0\#\rho_1} 
\end{equation}
so that $\rho_0$ and $\rho_1$ can be exchanged.
\end{remark}

\begin{remark}
Let $\rho_t=w_t^*w_t$  and $q$ be as in the proof of 
Proposition \ref{prop}. Then, in general,  $\rho_t$ is not the RLD geodesic. 
Indeed,  it can be easily computed that for the RLD metric, the geodesic 
equation reads
$$
\ddot{\rho}_t+\frac1{L_{\rho_t}+R_{\rho_t}}(\dot {\rho_t}^2)-
\dot{\rho_t}\rho_t^{-1}\dot{\rho_t}=a(t)\dot{\rho_t}
$$
where $a$ is a smooth function $a:I\to \mathbb{R}$, see also \cite{ditt00}.
We have 
$$
\rho_t=w_t^*w_t=\rho_0^{1/2}(1+t(q-1))^2\rho_0^{1/2}
$$ 
It can be
shown by direct computation that the geodesic equation is satisfied if and only
if 
$$
q(\rho_0q-q\rho_0)=(\rho_0q-q\rho_0)q
$$
which, for self-adjoint operators, implies
$q\rho_0=\rho_0q$. It follows that the inequality in Proposition \ref{prop} is
strict, unless $\rho_0$ and $\rho_1$ commute. In that case,  the geodesic
distances are the same for all monotone metrics.
\end{remark}

In \cite{peru}, a  class of generalized relative entropies
$$
H_g(\rho_0,\rho_1)=\Tr\rho_0g(\rho_0^{-1/2}\rho_1\rho_0^{-1}) 
$$
was introduced, here $g$ is an operator convex function.
This is a non-commutative version of the $f$-divergence. 
It was shown in \cite{peru} that the generalized entropy 
$H_g$ leads to a constant multiple of the RLD metric for infinitesimaly 
close elements in $\De$.

It is  easy to see  that
the right hand side of (\ref{eq:eq}) is equal to
$\sqrt{2H_{g_0}(\rho_0,\rho_1)}$,
where
\begin{equation}\label{eq:hnula}
g_0(t)=2+2t-4t^{1/2}. 
\end{equation}
Note that on commuting elements,
$H_{g_0}$ is equal to the Hellinger distance.

By  maximality of the RLD metric, we get

\begin{coro} Let  $\rho_0,\rho_1\in \Em$ and let $\lambda$ be a monotone metric. Then
$$
d_{Bures}(\rho_0,\rho_1)\le d_{\lambda}(\rho_0,\rho_1)\le
\sqrt{2H_{g_0}(\rho_0,\rho_1)}<
2\sqrt{\Tr\rho_0+\Tr\rho_1}
$$
If $\rho_0,\rho_1\in \De$, then 
$$
2\arccos \Tr(\rho_0^{1/2}\rho_1\rho_0^{1/2})^{1/2}\le D_{\lambda}(\rho_0,\rho_1)\le
2\arccos \Tr \rho_0\#\rho_1<\pi
$$
\end{coro}

\section{The wyd metrics}\label{sec:wyd}

The Wigner-Yanase-Dyson (WYD) metrics are defined by
$$
\lambda_{\rho}^{\alpha}(h,k)=\frac{\partial^2}{\partial t\partial s}
\Tr f_{\alpha}(\rho+th)f_{-\alpha}(\rho+sk)|_{s,t=0}
$$
where 
$$
f_{\alpha}(x)=\left\{ \begin{array}{lr} \frac2{1-\alpha}x^{\frac{1-\alpha}2} &
                                         \alpha\ne 1 \\ 
					\log(x) & \alpha=1
			\end{array}
		\right.	         
$$
As it was shown in \cite{hape}, these metrics are monotone for
$\alpha\in[-3,3]$. The family of WYD metrics is important in quantum
information geometry, see \cite{hase03, ja03,grasselli}.
As special cases, for  $\alpha=\pm 1$, we get the well known 
Bogoljubov-Kubo-Mori metric and  for $\alpha=\pm 3$ we get the RLD metric.

The smalest in this family is the Wigner-Yanase (WY) metric, obtained for
$\alpha=0$. The WY metric has the form
$$
\lambda_{\rho}(h,k)=4\Tr h (\sqrt{L_{\rho}}+\sqrt{R_{\rho}})^{-2}(k)
$$
The corresponding  geodesic path and distance was computed in
\cite{giis03}, using a non-commutative version of the square root 
map (\ref{eq:sqrt}) and a pull-back technique. 
We will show that these can be also
easily obtained using a similar method as in the Bures case.   

Let $\rho_t$ be a curve in $\Em$. Among its lifts $w_tw_t^*=\rho_t$, we will
again choose a horizontal one. In this case, the lift $w_t$ is horizontal if
it is contained in the natural positive cone at $w_0$, that is, if
$w_t=\rho_t^{1/2}u_0$ for all $t$. In this case,  the horizontal subspace is 
$H^0_w=\{ gu,\ g=g^*\}$, where $w=\rho^{1/2}u$ is the polar decomposition of
$w$. Each tangent vector $h\in T_{ww^*}\Em$ has a unique horizontal lift 
$h^0=gu\in H_w^0$, such that
$h=\Pi(h^0)=g\rho^{1/2}+\rho^{1/2}g$. The induced metric
$$
\lambda_{\rho}(h,k)=4\re\<h^0,k^0\>=4\Tr h(L_{\rho}^{1/2}+R_{\rho}^{1/2})^{2}(k)
$$
is the WY metric. Note that in this case $x\in H^0_w$ if and only
if $x^*\in H^0_{w^*}$, so that the WY metric is self-dual, in the sense
mentioned in Section \ref{sec:representation}. Let us also remark that it is
possible to obtain all the monotone metrics in a similar manner, see
\cite{diuh,ja02}.

Let now $\rho_0$ and $\rho_1$ be in $\Em$ and let $\rho_t$ be a curve
connecting them. Again, the WY lenght of $\rho_t$ is twice the Hilbert space 
length of its
horizontal lift $w_t=\rho_t^{1/2}u_0$. Therefore, $\rho_t$ is the 
geodesic path if  $w_t=t\rho_1^{1/2}u_0+(1-t)\rho_0^{1/2}u_0$, that is
$$
\rho_t=(t\rho_1^{1/2}+(1-t)\rho_0^{1/2})^2
$$
and the geodesic  distance is
$$
d_{WY}(\rho_0,\rho_1)=2\|\rho_0^{1/2}-\rho_1^{1/2}\|=2\sqrt{\Tr
\rho_0+\Tr\rho_1-2\Tr\rho_0^{1/2}\rho_1^{1/2}}
$$
Similarly, if $\rho_0,\rho_1\in \De$, then $\rho_t$ is a  geodesic path 
if and only if $w_t$ lies on the large circle connecting $\rho_0^{1/2}u_0$ and
$\rho_1^{1/2}u_0$. Hence
$$
\rho_t=\frac{(t\rho_1^{1/2}+(1-t)\rho_0^{1/2})^2}
{\|t\rho_1^{1/2}+(1-t)\rho_0^{1/2}\|^2}
$$
and
$$
D_{WY}(\rho_0,\rho_1)=2\arccos\Tr \rho_0^{1/2}\rho_1^{1/2}
$$

Let us denote by $\Delta_{\sigma,\rho}=L_{\sigma}R^{-1}_{\rho}$ the
relative modular operator. In \cite{petz86}, a class of quasi-entropies was
introduced by
$$
S_g(\rho,\sigma)=\Tr \rho^{1/2}g(\Delta_{\sigma,\rho})(\rho^{1/2})
$$
where $g$ is an operator convex function. This is another quantum version
of the $f$-divergences. It is easy to see that 
\begin{equation}\label{eq:dist}
d_{WY}(\rho_0,\rho_1)=\sqrt{2S_{g_0}(\rho_0,\rho_1)},
\end{equation}
where $g_0$ is   given by (\ref{eq:hnula}). It was proved in \cite{leru99} that
each monotone metric can be obtained as the hessian of $S_g$ for a suitable
operator convex function $g$. The choice $g=g_0$ leads to the WY metric.

From the previous Section and the fact that the WY metric 
is the least element in the family of WYD metrics, we get
\begin{coro} Let $\lambda$ be a WYD metric and $\rho_0,\rho_1\in \Em$. Then
$$
\sqrt{2S_{g_0}(\rho_0,\rho_1)}=d_{WY}(\rho_0,\rho_1)\le
d_{\lambda}(\rho_0,\rho_1)\le
\sqrt{2H_{g_0}(\rho_0,\rho_1)}
$$
where  $g_0(t)=2+2t-4t^{1/2}$. If $\rho_0,\rho_1\in \De$, then
$$
2\arccos\Tr\rho_0^{1/2}\rho_1^{1/2}\le D_{\lambda}(\rho_0,\rho_1)\le 2\arccos
\Tr\rho_0\#\rho_1
$$
\end{coro}

\section{Acknowledgements}

The research was supported by the grant VEGA 1/0264/03.

\end{document}